\documentclass[a4paper,12pt]{article}
\usepackage{psfig,epsfig}

\begin{document}

\title{The dielectric function excited by $\rho NN$ tensor coupling in nuclear matter}
\date{}
\author{Hui Liu \hspace{1.5cm}  Jia-rong Li \\\small\it email: liuhui@iopp.ccnu.edu.cn\\\small Institute of Particle Physics, Central China Normal University, \\\small Wuhan(430079), P.R.China}
\maketitle
\begin{abstract}
The dielectric function of nuclear matter excited by $\rho NN$ tensor coupling has been studied in the framework of Finite temperature field theory. The induced current mechanism has been introduced to explain the three extrema on the dielectric function curve, of which one is in the space-like region and the other two are in the time-like region. It points out that the tensor coupling contributes much more large amplitude than the vector coupling and plays a more important role on the time-like region compared with its effect on the space-like region.   

\end{abstract}

PACS number(s): 21.65.+f,11.10.Wx,13.75.Cs


\vspace{0.3cm}

The medium effects of the nuclear matter at finite temperature and/or density, such as the dispersion relation, the dielectric function (namely $\varepsilon$), the permeability and the nuclear effective mass, are very important facing the high energy physics today and excited a lot of investigations[1-8]. Among these, the dielectric property of strongly interacting matter (nuclear and quark matter) is a significant one, for it dominates the differences between the field in vacuum and in medium. Therefore all the medium effects in terms of the field in medium may be given by it in principle. For example, both the imaginary and real parts of $\varepsilon$ are firmly related some energy loss phenomenon such as jet quenching in nuclear reaction\cite{q,i}.

In relativistic heavy ion collisions, light vector mesons have achieved particular attention. This is because they can decay into dilepton pairs hence provide an almost undisturbed penetrating signal to study the properties of the fireball formed in the initial stage of the collisions\cite{r,s}. Among the light vector mesons, the $\rho$ meson acquires a special importance. Some theories and experimental results measured at CERN/SPS suggest that the in-medium properties of light vector mesons, especially the spectral function the $\rho$ meson, are crucial to understand the low mass enhancement of the lepton pairs\cite{h,o}. Besides this, due to its large decay width compared with $\omega$ and $\phi$'s, the $\rho$ meson has shorter life-time than that of the central fireball created in the relativistic heavy ion collisions, so as to be expected to carry more valuable information about the fireball.

The dielectric functions and the dispersion relations of the QED and QCD plasma with Hard thermal loop approximation have been discussed in some papers\cite{j}. The dielectric function excited by the  $\omega$ meson was mentioned in Ref. \cite{d} with the Hard nucleon loop approximation. And the dispersion relation of the $\rho$ meson in medium has also been studied\cite{c,f}. But the dielectric function of the $\rho$ meson is not yet clear and deserves further study. In the study of the properties of the $\rho$ meson, if only the vector coupling was involved in the interactive Lagrangian,
\begin{equation}
\mathcal{L}_{int}=g_{\rho NN}\bar\psi\gamma^\mu\tau_a\psi A_\mu^a,
\end{equation}
where $\psi$ is the nuclear field and $A_\mu^a$ is the $\rho$ meson field, some important information would have been lost. In view of the relatively large value of the iso-vector magnetic moment of the nucleon as compared to the iso-scalar case, the tensor coupling is more important for the $\rho$ meson than it is for other iso-scalar meson fields. In many experiments and phenomenological theories such viewpoint has been proved correct\cite{p,u}. The strong sensitivity of the $\rho$ meson mass to the $\rho NN$ tensor coupling have also been investigated by some authors\cite{b}. In order to carry on a quantitative calculation, the tensor coupling constant is included along with the vector coupling in the interactive Lagrangian: 
\begin{equation}
\mathcal{L}_{int}=g_{\rho NN}(\bar\psi\gamma^\mu\tau_a\psi A_\mu^a-\frac{\kappa_\rho}{2m_N}\bar\psi\sigma_{\mu\nu}\tau^a\psi\partial^\nu V_a^\mu),
\end{equation}  
where $\sigma_{\mu\nu}=\frac{i[\gamma_\mu,\gamma_\nu]}{2}$ and $m_N$ is the nucleon mass. In this paper, the dielectric function of nuclear matter excited by the $\rho NN$ coupling will be calculated with the above Lagrangian at finite temperature and chemical potential. In the following discussion we denote $K^2=K_0^2-\mathbf{k}^2$ and $|\mathbf{k}|=k$.


The expression of the dielectric function in terms of the polarization tensor has been defined through\cite{j,d}:

\begin{equation}
\varepsilon=1-\frac{\Pi_L}{K^2-m^2_\rho},
\end{equation}
where the polarization tensor can be decomposed into transverse and longitudinal parts as: $\Pi^{\mu\nu}=P^{\mu\nu}_L\Pi_L+P^{\mu\nu}_T\Pi_T$. The transverse and longitudinal projects are defined by Kapusta\cite{k}:

\begin{equation}
P^{00}_L=P^{0i}_L=P^{i0}_L=0, P^{ij}_L=\delta^{ij}-\frac{k^i k^j}{K^2},\  P^{\mu\nu}_T
=\frac{k^\mu k^\nu}{K^2}-g^{\mu\nu}-P^{\mu\nu}_T. 
\end{equation}


For the tensor coupling nucleon-polarization:

\begin{figure}[!ht]
 \begin{center}
   \scalebox{1}{\includegraphics*{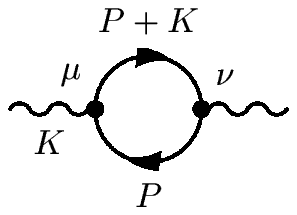}}
   \par{FIG. 1. The polarization tensor of $\rho NN$ tensor coupling}
 \end{center}
\end{figure}

\vspace{-1.5cm}

\begin{equation}
\Pi_{\mu\nu}^{\rho NN}=-2ig_{\rho NN}^2\int\frac{d^4P}{(2\pi)^4}Tr[\Gamma_\mu(K) G(P)\Gamma_\nu(-K) G(P+K)],
\end{equation}
where G(P) is the thermal nuclear propagator $G(P)=G_F(P)+G_D(P)$, and $\Gamma_\mu(K)=\gamma_\mu+\frac{i\kappa_\rho}{2m_N}\sigma_{\mu\nu}K^\nu$,

\begin{equation}
G_F(P)=\frac{\slash \kern-0.62em P+{m^*_N}}{P^2-{m^*_N}^2+i\epsilon},
\end{equation}

\begin{equation}
G_D(P)=2\pi i(P\!\!\!\!\slash+{m^*_N})n_F\delta(P^2-{m^*_N}^2) ,
\end{equation}
where $m_N^{*}$ is the effective mass of nucleon, and $n_F=[exp(\beta\omega_p-\mu)+1]^{-1}$ is Fermi-Dirac distribution.
\begin{equation}
\Pi_{\mu\nu}(K)=\Pi_{\mu\nu}^F(K)+\Pi_{\mu\nu}^D(K),
\end{equation}

\begin{eqnarray}
\Pi^{F}_{\mu\nu}(K)&=&-2ig_{\rho NN}^2\int\frac{d^4P}{(2\pi)^4}Tr[\Gamma_\mu(K) G_F(P)\Gamma_\nu(-K) G_F(P+K)], \\
\Pi^{D}_{\mu\nu}(K)&=&-2ig_{\rho NN}^2\int\frac{d^4P}{(2\pi)^4}Tr[\Gamma_\mu(K) G_F(P)\gamma_\nu(-K) G_D(P+K),\nonumber\\
+&&\hskip -1cm \Gamma_\mu(K) G_D(P)\Gamma_\nu(-K) G_F(P+K)+\Gamma_\mu(K) G_D(P)\Gamma_\nu(-K) G_D(P+K)].\nonumber \\
\end{eqnarray}
The Feynman part is divergent but can be renormalized\cite{p}. Here we only focus on the temperature-dependent part. 

By using Eq. (4), it's easy to get $\Pi^L=\frac{K^2}{k^2}\Pi_{00}$. And with tedious calculation one can obtain 

\begin{equation}
\Pi^{D,L}_{\rho NN}=\Pi^L_1+2\Pi^L_2+\Pi^L_3,
\end{equation}
\begin{eqnarray}
\Pi^L_1&=&\frac{g_{\rho NN}^2}{\pi^2}\frac{K^2}{k^2}\int\frac{p^2dp}{\omega_p}[\frac{4\omega_p^2-4\omega_pK_0+K^2}{4pk}\log{A} \nonumber \\
&   &+\frac{4\omega_p^2+4\omega_p K_0+K^2}{4pk}\log{B}-2](n_F+\bar n_F),\\ 
\Pi^L_2&=&-\frac{g_{\rho NN}^2}{\pi^2}\frac{K^2}{k^2}\frac{\kappa_\rho}{2m_N}\int\frac{pdp}{\omega_p}[\log{A}+\log{B}](n_F+\bar n_F),\\
\Pi^L_3&=&-\frac{g_{\rho NN}^2}{4\pi^2}\frac{K^2}{k^2}(\frac{\kappa_\rho}{2m_N})^2\int\frac{p^2dp}{\omega_p}[\frac{k^2(K_2-4p^2)+(K^2-2\omega K_0)^2}{pk}\log{A}\nonumber \\
&   &+\frac{k^2(K_2-4p^2)+(K^2+2\omega K_0)^2}{pk}\log{B}-8K_0^2](n_F+\bar n_F),
\end{eqnarray}
here, $\Pi^L_1$ is the result of the vector coupling.
The parameters are defined by 
\begin{eqnarray}
\omega_p=\sqrt{p^2+m_N^{*2}}, \ \ \ \bar n_F=[exp(\beta\omega_p+\mu)+1]^{-1},\nonumber
\end{eqnarray}
\begin{equation}
A=\frac{-2\omega_p K_0+K^2+2pk}{-2\omega_pK_0+K^2-2pk} , B=\frac{2\omega_pK_0+K^2+2pk}{2\omega_pK_0+K^2-2pk}.
\end{equation}

Joining up Eq. (3) and Eq. (11), and taking the real part of the dielectric function into account, one can display the curves in Fig. 2. In the numerical calculation, the effective mass of nucleon varies with $\mu$ and T which is decided by the self-consistent equation\cite{l}. The other parameters are in TABLE I.

TABLE I. Parameters in numerical calculation\cite{l}. The tensor coupling constant are adjusted to fit some experimental data phenomenologically\cite{p,u}. The masses, the temperature and the chemical potential are in (MeV).

\begin{center}
\begin{tabular}{c|c|c|c|c|c|c|c|c}
\hline\hline
       & $m_N^*(\mu=800)$ & $m_N^*(\mu=850)$ &$\kappa$ &  $g_\rho^2$ & $g_s^2$  & $m_\rho$ & $m_s$ & $m_N$ \\
\hline
T=10   & 767.7            & 830.8            &   6.1    &  6.91      & 62.89    & 770      & 550  & 939    \\
\cline{1-3}
T=20   & 780.3           &  850.5           &            &            &          &        &       &       \\
\hline\hline
\end{tabular}
\end{center}

\begin{figure}[!ht]
 \begin{center}
   \resizebox{10cm}{!}{\includegraphics*{fig2.epsi}}
   \par{FIG. 2. The dielectric function of $\rho NN$ tensor coupling in terms of $K_0/k$. The solid line is for the case of T=10MeV, $\mu$=800MeV, the dashed line for T=10MeV,$\mu$=850MeV, and the dotted line for T=20MeV, $\mu$=800MeV.}
 \end{center}
\end{figure}

\begin{figure}[!ht]
 \begin{center}
   \resizebox{10cm}{!}{\includegraphics*{fig3.epsi}}
   \par{FIG. 3. The exaggeration of FIG. 2 in $0\sim 2$ region.}
 \end{center}
\end{figure}

In Fig. 2,  $k$ has been fixed to 100MeV as to see the dielectric function varying with frequency. The solid curve contains one singularity at $K_0/k \approx 7.8$ and two extrema (where $K_0/k \approx 16$) with their positions and amplitude changing with $\mu$ and T. These two extrema appear at nearly the same point in the time-like region. On one hand their amplitudes increase when  $\mu$ or T rises, and specially one should notice the detail that these two extrema in the time-like region become more and more unsymmetricalwith respect to the transverse axis when T is rising. This will be explained physically in the following. On the other hand, the change of the chemical potential moves the position of the extrema, and has a tiny effect on amplitude.

Generally speaking, it is impossible for a in-medium meson to meet the mass-shell condition  $K_0^2-k^2=m_\rho^2$, which brings the singularity (See Eq. (3)). So the singularity might be trivial to some extent. What we are interested in is the two unsymmetrical extrema. As we know that the polarization is the essence of the induced current in medium, which is  explained clearly by the formula,

\begin{equation}
j^\mu_{ind}(x)=\int d^4x'\Pi^{\mu\nu}(x-x')A_\nu(x')
\end{equation} 
where $A_\nu(x')$ is the total field and $\Pi^{\mu\nu}(x-x')$ is the polarization tensor, and note that the fourth component of $j^\mu_{ind}(x)$ is the induced charge density, one may immediately recognize that the induced current determines the dielectric function directly in the analogy of the electromagnetism. Detailed analyse to the numerical result shown in Fig. 2, one may discover that the two extrema in the time-like region appear when the frequency $K_0\ge 2m_N^*$. For example, on the solid curve of Fig. 2, the position of the two extrema is around $K_0/k=16$, i. e. $K_0=1600MeV$, and $2m_N^*\sim 1500MeV$. If we take the induced current mechanism into account, this phenomenon will be explained naturally. When the energy of the meson ($K_0$) is large enough to cross over the threshold of $2m_N^*$, the nucleon will be excited from the Fermi sea, leaving a hole simultaneously. These excited nucleons and holes will form the polarized current in medium and then make the $\varepsilon$ reach its extrema. This explanation on the induced current mechanism implies that the energy of meson has been absorbed by the nucleon . As the imaginary part of the dielectric function reflects the energy exchange, it is wise to check the result from this point of view. Fortunately we found an extremum on the imaginary part in the time-like region, whose position was exactly the same with that on the real part curve. And what is more interesting is that besides this extremum in the time-like region for the imaginary part, there exists another unexpected one in the space-like region. This strange 'additional' extremum lead us to seek something unusual on the real part curve. Exaggerating the region of $0\sim1$ of Fig. 2, one can certainly find an relevant extremum on the real part curve, where  $K_0/k \approx 0.3$ (Fig. 3). It is obviously that this extremum in the space-like region cannot be explained by the above induced current mechanism, for here $K^2=K_0^2-k^2<0$. In fact this small extremum can be explained by another induced current mechanism. But note that the space-like region is the Landau damping area. In this area, the meson, as an excitation mode, whose phase velocity $K_0/k<1$ (i. e. less than the light velocity), will loss its energy into the nucleons on the Fermi face (i. e. the excitation mode is damped down) and these excited necleons then form a corresponding induced current, which causes the deviation of the dielectric function from its value in vacuum and contributes the extremum in the space-like region. Therefore people usually name it as the Landau damping mechanism.

So far we have obtained one singularity and three extrema on the real part of the dielectric function curve. Let us make a summary. As what is shown in Fig. 2 accompanied by Fig. 3, in the space-like region, $\varepsilon$ decreases first with the increase of $K_0$ and displays a minimum value, which has very rather amplitude and reflects the Landau damping effect. Then with $K_0$ rising, the curve enters into the time-like region, and $\varepsilon$ rises until the singularity arises when the mass-shell condition of physical particle $K_0^2-k^2=m_\rho^2$ is satisfied. After crossing the singularity $\varepsilon$ is rising again with $K_0$. But when the frequency $K_0 \ge 2m_N^*$, the nucleons begin to be excited from the Fermi sea, and form the induced current which leads the $\varepsilon$ deviating far from its vacuum value 1. The curve gives another two extrema with almost symmetrical structure. One of those extrema is contributed by excited nucleons and the other is attributed to holes. At low temperature, the number of nucleons and holes is almost in balance. But when T reaches a certain degree, the fermi energy level will be destroyed. At that time, the nuleons are more than holes, thus they will have unsymmetrical contribution to the dielectric function. This is the reason that when T is rising, more and more unsymmetrical structure of the two extrema in the time-like region is demonstrated in Fig. 2.

Finally, let us compare the tensor coupling with the vector coupling. As it is pointed out that the Eq. (12) is the polarization tensor of the vector coupling, whose relevant dielectric function has been plotted in Fig. 4. One may find that the dielectric function of tensor coupling has totally different order of magnitude with that of the vector coupling. And the Figs. 2 and 3 show that the magnitudes of the extrema in the time-like region are much larger than the one in the space-like region. This result tells us that the tensor coupling plays a more important role in the time-like region than its effect on the space-like region in the nuclear medium.  


\begin{figure}[!ht]
 \begin{center}
   \resizebox{10cm}{!}{\includegraphics*{fig4.epsi}}
   \par{FIG. 4. The dielectric function excited by $\rho NN$ vector coupling with  T=10MeV, $\mu$=800MeV.}
 \end{center}
\end{figure}

Let us draw some conclusions on dielectric function of nuclear matter excited by $\rho NN$ tensor coupling. Firstly, the dielectric function appears singularity when the mass-shell condition is satisfied. Besides the singularity, it has three extrema, one is in the space-like region and the other two, appearing at almost the same point, in the time-like region. Secondly, the extrema are explained physically by the corresponding induced current mechanism. Especially, the unsymmetrical structure of the two extrema in the  time-like region are related to the slightly destroyed Fermi energy level at finite temperature. Finally, the tensor coupling dominates the contribution to $\varepsilon$ compared with vector coupling and is more important in the time-like region.

\vspace{0.3cm}
\begin{center}
\bf ACKNOWLEDGEMENTS
\end{center}

This work is supported by the National Natural Science Foundation of China, No. 10175026 and No. 90303007.



\begin{thebibliography}{aaaaaaaaaa}
\bibitem{t} G.E. Brown and M. Rho, Phys. Rev. Lett. 66, 2720 (1991)
\bibitem{a} C. Song, P.W. Xia and C.M. Ko, Phys. Rev. C52, 408 (1995)
\bibitem{g} O. Teodorescu, A.K. Dutt-Mazumder, C. Gale, Phys. Rev. C66, 015209 (2002)
\bibitem{c} Ji-sheng Chen, Jia-rong Li and Peng-fei Zhuang, Phys. Rev. C67, 068202 (2003)
\bibitem{f} W. Peters, M. Post, S. Leupold and U. Mosel, Nucl. Phys A632, 109 (1998)
\bibitem{e} K. Saito and A.W. Thomas, Phys. Rev. C51, 2757 (1995)
\bibitem{h} A. Mishra, J. Reinhardt, H. Stocker, W. Greiner, Phys. Rev. C66, 064902 (2002)
\bibitem{b} A. Mishra, J. C. Parikh, W. Greiner, J. Phys. G28, 151 (2002)
\bibitem{q} M.H. Thoma, J.Phys. G26, 1507 (2000) 
\bibitem{i} M.H. Thoma and M. Gylassy Nucl.Phys. B351, 491 (1991) 
\bibitem{r} K. Kajantie, J. Kapusta, L. McLerran and A. Mekjian, Phys. Rev. D34, 2746 (1986) 
\bibitem{s} I. Tserruya, Nucl. Phys. A590, 127c (1995) 
\bibitem{o} Ji-sheng Chen, Jia-rong Li, and Peng-fei Zhuang, JHEP 0211, 014(2002)
\bibitem{j} H.A. Weldon, Phys.Rev. D26, 1394 (1982)
\bibitem{d} A.K. Dutt-Mazumder, Nucl. Phys. A713, 119 (2003)
\bibitem{p} H. Shiomi and T.Hatsuda, Phys. Lett. B334, 281 (1994)
\bibitem{u} R. Machaleit, in Advances in Nuclear Physics, edit by J. W. Negele and E. Vogt (Plenum, New York,             1986), Vol. 19
\bibitem{k} J.I. Kapusta, Finite-Temperature Field Theory (Cambridge Univ. Press 1989)
\bibitem{l} B.D. Serot and J.D. Walecka Adv. Nucl. Phys. V16 (1986)


\end{thebibliography}
\end{document}